\documentclass[superscriptaddress,floatfix,twocolumn, showpacs,preprintnumbers,amsmath,amssymb,aps, prl, longbibliography]{revtex4-1}

\usepackage{sidecap} 

\usepackage{graphicx}
\usepackage{subfigure}
\usepackage{dcolumn}
\usepackage{bm}

\begin{document}

\title{Spectro-temporal shaping of seeded free-electron laser pulses}

\author{David Gauthier}
\author{Primo\v{z} Rebernik Ribi\v{c}}
\affiliation{Elettra-Sincrotrone Trieste, 34149 Trieste, Italy}
\author{Giovanni De Ninno}
\affiliation{Elettra-Sincrotrone Trieste, 34149 Trieste, Italy}
\affiliation{Laboratory of Quantum Optics, University of Nova Gorica, 5001 Nova Gorica, Slovenia}
\author{Enrico Allaria}
\affiliation{Elettra-Sincrotrone Trieste, 34149 Trieste, Italy}
\author{Paolo Cinquegrana}
\affiliation{Elettra-Sincrotrone Trieste, 34149 Trieste, Italy}
\author{Miltcho Boyanov Danailov}
\affiliation{Elettra-Sincrotrone Trieste, 34149 Trieste, Italy}
\author{Alexander Demidovich}
\affiliation{Elettra-Sincrotrone Trieste, 34149 Trieste, Italy}
\author{Eugenio Ferrari}
\affiliation{Elettra-Sincrotrone Trieste, 34149 Trieste, Italy}
\affiliation{Universit\`{a} degli Studi di Trieste, Dipartimento di Fisica, Piazzale Europa 1, 34100 Trieste, Italy}
\author{Luca Giannessi}
\affiliation{Elettra-Sincrotrone Trieste, 34149 Trieste, Italy}
\affiliation{Theory Group ENEA C.R. Frascati, Via E. Fermi 45, 00044 Frascati, Italy}
\author{Beno\^{i}t Mahieu}
\affiliation{Laboratoire d'Optique Appliqu\'{e}e, UMR 7639, ENSTA-CNRS-\'{E}cole Polytechnique, Chemin de la Huni\`{e}re, 91761 Palaiseau, France}
\author{Giuseppe Penco}
\affiliation{Elettra-Sincrotrone Trieste, 34149 Trieste, Italy}

\date{\today}

\begin{abstract}
We demonstrate the ability to control and shape the spectro-temporal content of extreme-ultraviolet (XUV) pulses produced by a seeded free-electron laser (FEL). The control over the spectro-temporal properties of XUV light was achieved by precisely manipulating the linear frequency chirp of the seed laser. Our results agree with existing theory, which allows retrieving the temporal properties (amplitude and phase) of the FEL pulse from measurements of the spectra as a function of the FEL operating parameters. Furthermore, we show the first direct evidence of the full temporal coherence of FEL light and generate Fourier limited pulses by fine-tuning the FEL temporal phase. The possibility to tailor the spectro-temporal content of intense short-wavelength pulses represents the first step towards efficient nonlinear optics in the XUV to X-ray spectral region and will enable precise manipulation of core-electron excitations using the methods of coherent quantum control.

\end{abstract}

\maketitle

\section{}

The development of the first high-power pulsed lasers in the 1960s marks an important milestone in the long-standing effort to actively control the temporal evolution of quantum-mechanical systems. However, it was not until the early 1990s, when ultrashort pulse shaping techniques became practical, that coherent control of quantum phenomena finally became a reality \cite{warren:1993}. Tailoring the spectro-temporal content of intense laser pulses in the visible range opened up countless possibilities for manipulating the quantum state of matter, with examples ranging from control of population transfer in optical transitions \cite{chen:1990} and currents in semiconductors \cite{dupont:1995} to control of chemical reactions \cite{potter:1992} and energy flow in biomolecular complexes \cite{herek:2002}, and many others (see, e.g. Ref. \cite{brif:2010}  and references therein). Undoubtedly, the ability to shape the spectro-temporal content of powerful extreme-ultraviolet (XUV) and X-ray pulses would trigger widespread efforts to extend the concepts of coherent control into the short-wavelength regime, leading ultimately to the development of new methods for probing and manipulating core electrons in atoms, molecules and materials. As a more immediate application, it may find use in numerous advanced spectroscopic techniques such as resonant inelastic X-ray scattering \cite{luuk:2011} and coherent ultrafast core-hole correlation spectroscopy, offering unique capabilities for probing elementary excitations \cite{schweigert:2007}.

In the XUV/X-ray spectral region, free-electron lasers (FELs) are currently the only devices that can deliver femtosecond laser-like pulses with peak powers in the gigawatt range \cite{ackermann:2007,emma:2010,allaria:2012,ishikawa:2012,allaria:2013}. However, the ability to generate fully coherent pulses and to shape their spectro-temporal content with high stability on a shot-to-shot basis is extremely challenging, due to the difficulties in precisely controlling the light generation process. In this Letter we show that the spectro-temporal content of powerful ultrashort XUV pulses can be precisely shaped using a laser-seeded FEL. By tuning the seed laser operating parameters we generate intense femtosecond XUV pulses with a controllable amount of linear frequency chirp, which can be ultimately reduced to zero. Our results constitute the first experimental evidence of Fourier limited pulses from an XUV FEL and pave the way to full spectro-temporal shaping of powerful ultrafast pulses in the XUV to soft-X-ray spectral region.

\begin{figure*}
\includegraphics[width=0.99\textwidth]{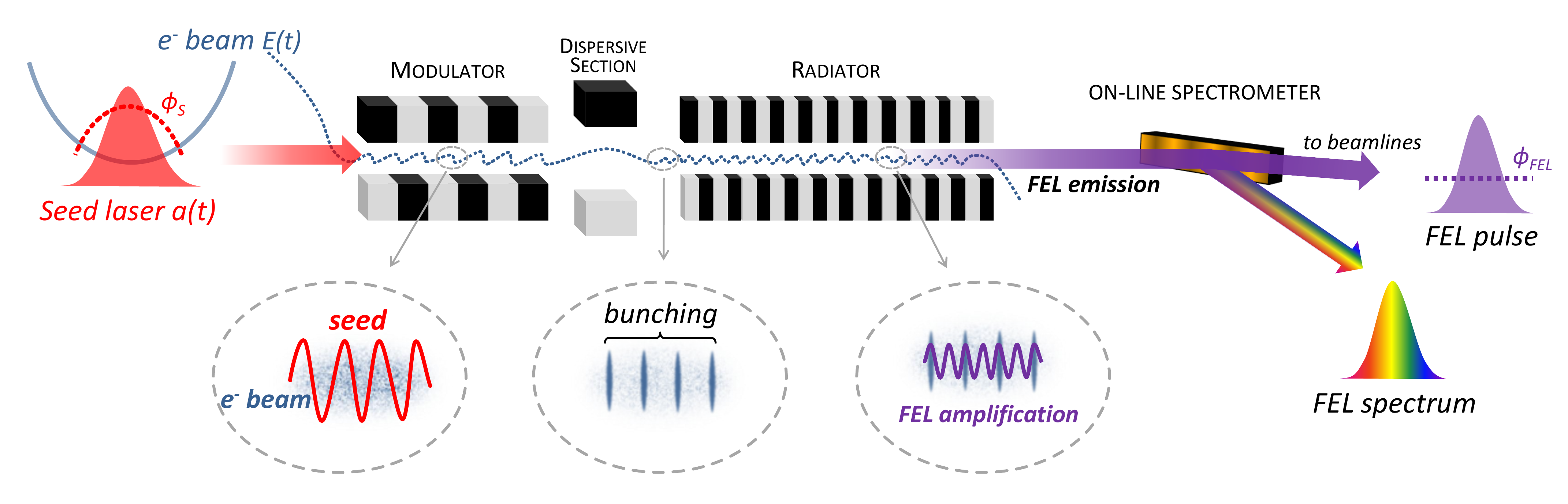} 
  \caption{The high-gain harmonic generation scheme. After interacting with the seed laser in the first undulator (modulator) and with the magnetic field of the dispersive section, the relativistic e-beam develops periodic density modulations (bunching) at harmonics of the seed wavelength. The microbunched e-beam is then injected into the second undulator (radiator), whose periodic magnetic field forces the electrons to emit coherent and powerful light pulses in the XUV or soft-X-ray region. The spectro-temporal properties of radiation are controlled by manipulating the seed laser electric field.}
  \label{hhgscheme}
\end{figure*}

A natural approach towards fully controlling the spectro-temporal profile of powerful XUV/X-ray pulses is to exploit the unique capabilities of a seeded FEL. Whereas in FELs based on SASE (self-amplified spontaneous emission) the amplification is triggered by spontaneous undulator emission \cite{bonifacio:1984}, a seeded FEL relies on a coherent input signal. We employed the high-gain harmonic generation (HGHG) seeding scheme \cite{yu:1991} available at the FERMI FEL \cite{allaria:2012}. In HGHG, see Fig. \ref{hhgscheme}, a seed laser is used to imprint a periodic energy modulation at the seed wavelength $\lambda$ (typically in the UV) onto a relativistic electron beam (e-beam) in the modulator. This energy modulation is converted into a current density modulation when the e-beam interacts with the magnetic field of the dispersive section. Such a microbunched e-beam emits coherent light at $\lambda/n$  ($n$ integer) in the XUV/X-ray region as the electrons traverse the periodic magnetic field of the radiator.

\newpage

The FEL output at the $n$th harmonic is driven by the electron bunching factor \cite{yu:1991,stupakov:2011,ratner:2012}
\begin{equation}
b_n(t)=e^{-n^2 B^2/2} J_n[-n B A(t)] e^{i n[\phi_s(t)+\phi_e(t)]} \mbox{,}
\label{bunching_factor}
\end{equation}
where $B$ is the dimensionless strength of the dispersive section (dispersive strength) \cite{stupakov:2011}, $J_n$ is the $n$th order Bessel function, and $A(t)$ is the time-dependent energy modulation (normalized to the electron energy spread $\sigma_E$), which is proportional to the envelope $a_0(t)$ of the seed laser electric field $a(t)=a_0(t)\sin{[\omega_0 t+\phi_s(t)]}$, $\omega_0$ being the central frequency. The exponent in the last factor, where $\phi_e(t)=(B/\sigma_E)E(t)$, accounts for the slowly varying phase of the seed $\phi_s(t)$ and a possible time-dependent energy profile $E(t)$ imprinted onto the e-beam by the linear accelerator (linac), which produces the relativistic electrons.

As the microbunched e-beam is injected into the radiator it starts emitting coherent light. Initially the e-beam is rigid and the FEL electric field is directly proportional to the bunching factor in Eq. \ref{bunching_factor}. In the linear regime (before saturation) \cite{bonifacio:1984,saldin:2010,dattoli:2013}, the field envelope is preserved despite amplification in the radiator, as demonstrated in the following. However, a small additional phase $\phi_a(t)$ is introduced \cite{wu:2007, lutman:2009,marinelli:2010}, so that the total FEL phase becomes:
\begin{equation}
\phi(t)_{FEL}=n[\phi_s(t)+\phi_e(t)]+\phi_a(t) \mbox{.}
\label{FEL_phase}
\end{equation}
The above equations provide the basis for FEL pulse shaping through the manipulation of the seed envelope $a_0(t)$ and phase $\phi_s(t)$.

Both the e-beam time-dependent energy profile and the phase developed during amplification affect the pulse properties \cite{stupakov:2011,wu:2007,lutman:2009,marinelli:2010}. However, as we show in the following, these effect can be fully compensated for by properly tuning the temporal phase of the seed laser. For the cases considered here it suffices to expand each of the individual phase contributions $\phi_s(t)$, $\phi_e(t)$ and $\phi_a(t)$ into a power series in time up to the second order. While a linear-term coefficient $d\phi(t)/dt$ results in an absolute frequency (wavelength) shift \cite{shaftan:2005}, a quadratic-term coefficient $d^2\phi(t)/dt^2$, called the chirp rate, gives rise to a linear frequency chirp in the pulse \cite{ratner:2012}. A suitable seed laser chirp can then be used to counter the combined effects due to the e-beam quadratic energy curvature and the chirp developed during amplification.

\begin{figure*}
\includegraphics[width=0.99\textwidth]{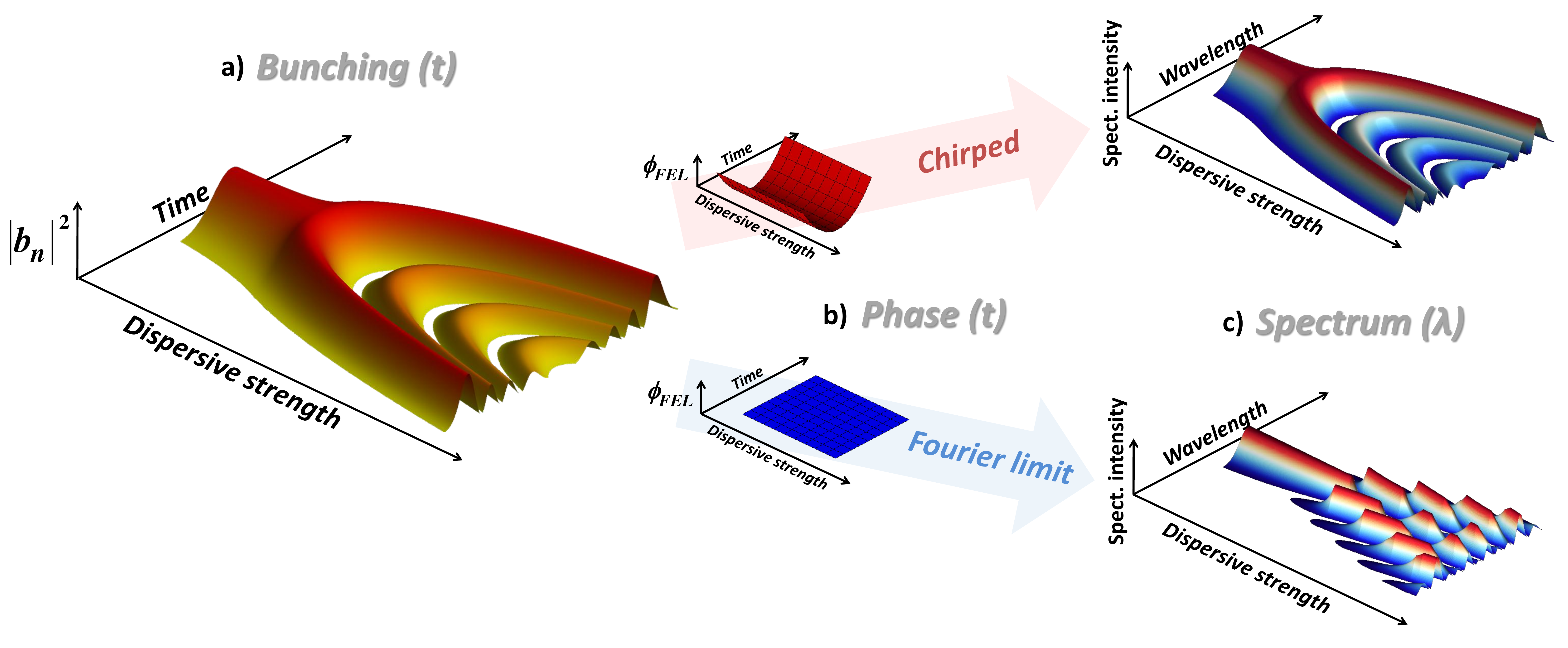} 
  \caption{FEL pulse shaping through manipulations of the electron bunching envelope and FEL phase. a), Theoretical bunching envelope (square modulus of $b_n(t)$, Eq. \ref{bunching_factor}), corresponding to the temporal FEL pulse shape, as a function of the dimensionless dispersive strength ($B$). Just before the bifurcation $B$ is optimized for maximum bunching, resulting in a single peak. Increasing $B$ leads to peak splitting due to electron overbunching in the central part. The envelope develops a multi-peak structure as a result of electron rebunching when $B$ is increased even further. c), The spectral map (FEL spectrum as a function of $B$) strongly depends on the FEL phase (b)). For a significantly chirped FEL pulse there is a direct correspondence between the temporal and spectral domains (top). On the other hand, the spectral map of a Fourier limited pulse (with a flat phase) develops distinctive features with increasing $B$ due to interference  between the individual peaks in the multi-peak bunching structure (bottom). For the sake of visualization, the bunching and spectral maps are normalized in amplitude for each value of $B$.}
\label{bunching_factor_figure}
\end{figure*}

Fig. \ref{bunching_factor_figure} highlights how the spectro-temporal content of FEL light can be shaped by tuning the operating parameters. First, Fig. \ref{bunching_factor_figure} a), the bunching envelope, generated here by a Gaussian seed, can be modified by changing the strength $B$ of the dispersive section. With increasing $B$, the bunching develops modulations as a function of time (Eq. \ref{bunching_factor}) due to the process of electron overbunching and rebunching \cite{labat:2009,xiang:2014}, leading to a pulsed structure. As emphasized above, the FEL pulse temporal shape directly corresponds to the bunching envelope. Second, manipulating the FEL temporal phase using a chirped seed, Fig. \ref{bunching_factor_figure}b), leads to a drastic modification of the FEL spectral content; see Fig. \ref{bunching_factor_figure} c). While the spectral map (spectrum vs. $B$) of a significantly chirped FEL pulse directly corresponds to its temporal map \cite{ninno:2013,gauthier:2013}, a Fourier limited pulse with a flat temporal phase shows a distinctively different spectral signature.

\begin{figure}
\includegraphics[width=0.40\textwidth]{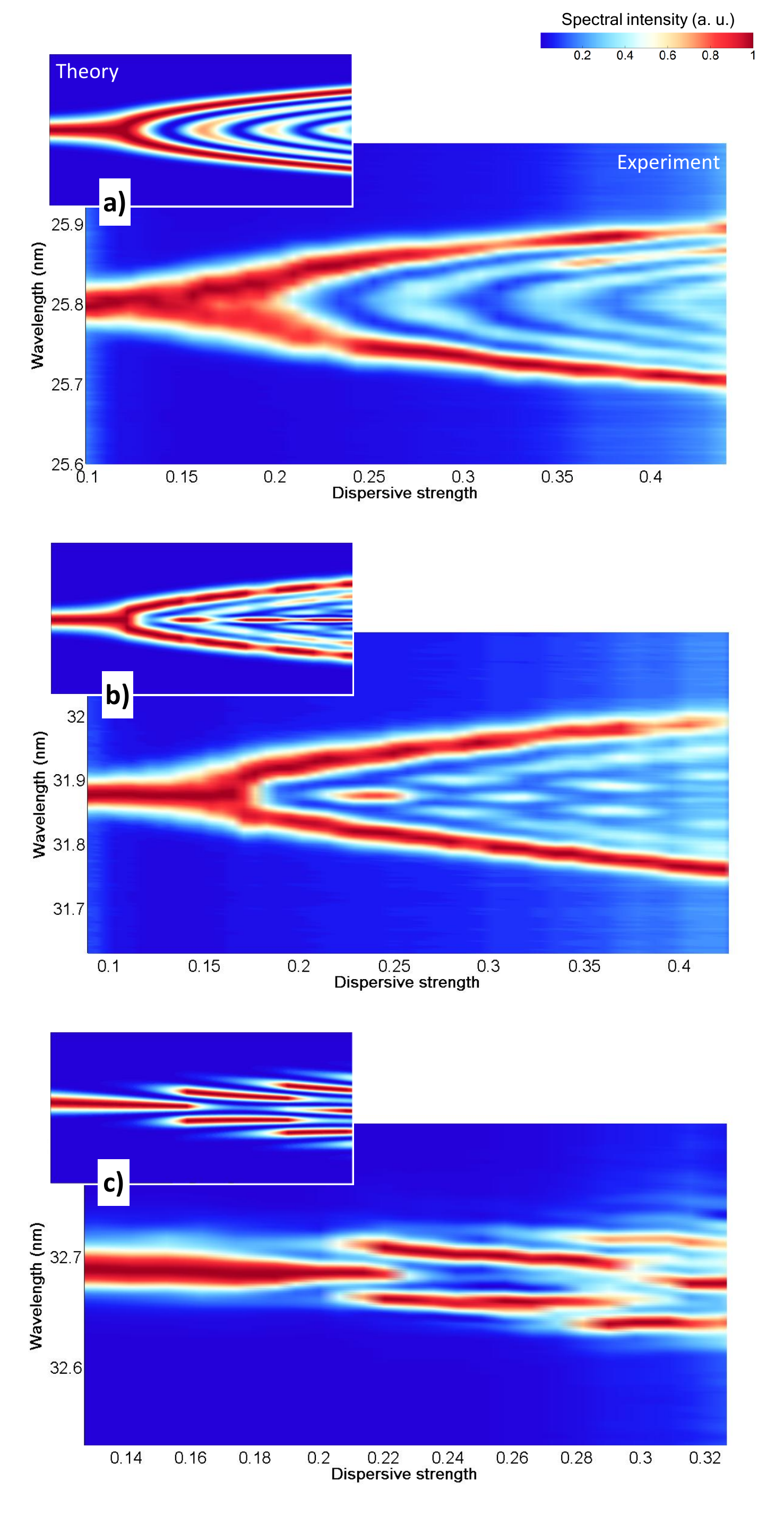} 
  \caption{FEL pulse spectra as a function of the dimensionless dispersive strength for three different conditions of the linear frequency chirp on the seed laser. a), Significant positive chirp; b), moderate positive chirp; and c), slight negative chirp for chirp compensated FEL pulses. Insets are theoretical spectral maps reproduced using Eqs. \ref{bunching_factor} and \ref{FEL_phase} in the spectral domain. The wavelength shift (spectral map tilting) in c) is due to the contribution of a linear-term coefficient $d\phi(t)/dt$ from the e-beam energy profile. The experiment in a) was done at $n=10$ using all of the six 2.4-m-long radiator sections, while for b) and c) $n=8$ and the number of active radiator sections was 4 and 5, respectively. In all cases, the e-beam energy, energy spread, current, bunch charge, and emittance were $\simeq 1.3$ GeV, $\simeq 150$ keV, $\simeq 600$ A, $\simeq 500$ pC, and $\simeq 1$ mm mrad, respectively. For a) and b) the e-beam (positive) quadratic energy curvature  was $\simeq$ 5 MeV/ps$^2$, while for c) the value was $\simeq$ 12 MeV/ps$^2$. The data was acquired with a dedicated spectrometer \cite{svetina:2011,zangrando:2015}. The maps in a) and b) consist of 10 accumulated single-shot spectra for each value of $B$, while data in c)  were taken by summing 3 multi-shot spectra (10 shots). The data within each spectral slice (for fixed $B$) are normalized in amplitude for better visualization.
}
\label{spectra}
\end{figure}

In the following we experimentally verify the above predictions and demonstrate the power of the HGHG scheme in controlling the spectro-temporal content of intense XUV pulses. Figure \ref{spectra} a) shows the experimental spectral map, when the relatively high positive frequency chirp on the FEL, dominated by the positive chirp due to a stretched seed, dictates the radiation properties. Here, we use the group delay dispersion (GDD) as a measure for the degree of chirp on the seed laser. The strong linear frequency chirp, characterized by a GDD of 7390 fs$^2$, was achieved by putting a 20 mm thick plate of fused silica into the seed laser (frequency up-converted pulses from an optical parametric amplifier, OPA) path, inducing a positive chirp rate of $5.9 \times 10^{-5}$ rad/fs$^2$ and stretching the seed pulse duration to 250 fs (FWHM), at the operating wavelength of 258 nm. The situation corresponds to the one in the top part of Fig. \ref{bunching_factor_figure}c). The FEL spectrum develops intensity modulations with increasing $B$, which directly correspond to the intensity modulations in the temporal domain. Excellent agreement \cite{rebunching} between experiment and theory (inset) \cite{theory} demonstrates that, despite amplification in the radiator, the FEL pulse envelope is preserved, justifying the use of Eqs. \ref{bunching_factor} and \ref{FEL_phase} to describe the spectro-temporal content of FEL light.

The next experiment, Fig. \ref{spectra}b), was performed with a moderate positive chirp (GDD = 4520 fs$^2$) on the seed laser. For this purpose the direct output of the OPA at 255 nm, with a pulse duration of 160 fs and a chirp rate of $6.7 \times 10^{-5}$ rad/fs$^2$, was used to seed the FEL. Because the total FEL chirp is not high enough to satisfy the condition for the spectro-temporal equivalence \cite{gauthier:2013}, the central part of the spectral map is significantly modified.  Again, a remarkably good correspondence is obtained between experiment and theory (inset), demonstrating the predictive power of Eqs. \ref{bunching_factor} and \ref{FEL_phase}.

Based on the two spectral signatures in Figs. \ref{spectra}a) and b) the next set of experiments was carried out by putting a negative chirp rate on the seed (third harmonic of a Ti:Sapphire laser, operating at 261.6 nm); in this case the GDD was -1100 fs$^2$. A fine control of the negative chirp rate was possible using an optical compressor based on a pair of gratings. The data in Fig. \ref{spectra}c) correspond to a chirp rate of $-2.0 \times 10^{-5}$ rad/fs$^2$, compressing the seed pulse down to 120 fs. A strong modification of the spectral content vs. $B$ with respect to the previous two cases is seen in Fig. \ref{spectra} c), confirming the extremely high sensitivity of the spectral maps to the FEL phase. Experimental data once again fit well with calculations (inset). Remarkably, the spectral signature corresponds to that of a Fourier limited pulse with a flat temporal phase (cf. Fig. \ref{bunching_factor_figure}c), bottom). The negative chirp rate on the seed compensates the positive chirp due to the quadratic curvature ($\simeq 12$ MeV/ps$^2$, measured at the end of the linac using a radio-frequency deflecting cavity in combination with an energy spectrometer \cite{craievich:2015}) on the e-beam time-dependent energy profile and the chirp developed during the amplification stage \cite{theory}. More precisely, because the phase contribution from the e-beam $\phi_e(t)$ is a function of $B$, the FEL chirp rate varies linearly from about $-1$ to $9 \times 10^{-5}$ rad/fs$^2$ in the range of dispersive strengths used in Fig. \ref{spectra} c), going through zero at $B\simeq 0.14$, where full chirp compensation is achieved.

Several important conclusions can be drawn from the above results. The experimental spectral maps in Fig. \ref{spectra} and their excellent agreement with theory based on Eqs. \ref{bunching_factor} and \ref{FEL_phase} imply that the FEL output can be fully manipulated by controlling the seed laser envelope and phase, where the dispersive strength can act as an additional tuning knob (e.g., an FEL pulse train with a fixed phase relationship between individual pulses is obtained simply by increasing the strength of the dispersive section). Our results demonstrate, to the best of our knowledge, the first compelling evidence that the radiation produced by a seeded FEL is temporally fully coherent and that by adjusting the seed laser phase, a controllable amount of linear frequency chirp can be transferred to the FEL pulse. Moreover, a careful tuning allows generation of Fourier limited pulses with a flat temporal phase. Such pulses are typically used as references in the field of coherent quantum control. Our results therefore open the door to full spectro-temporal shaping of intense ultrashort pulses in the XUV to soft-X-ray region using methods similar to the ones developed in the visible spectral region \cite{weiner:2011}. In combination with the possibility to engineer the FEL transverse radiation profile using HGHG \cite{ribic:2014}, this sets the stage for entirely new experiments with pulsed light at short wavelengths.

\section{Acknowledgments}
The authors acknowledge the assistance of the entire FERMI team during the preparation and tuning of the FEL and also wish to thank C. Svetina, M. Zangrando and the PADRES group for the support on the FERMI photon diagnostics. P.R.R. acknowledges financial support from the TALENTS UP Programme (7th FP, Specific Programme PEOPLE, Marie Curie Actions - COFUND) G.A. 600204. B.M. acknowledges financial support from the LabEx PALM (RAMSES project). The authors also acknowledge useful discussions with G. Stupakov and K. Prince.


\begin{thebibliography}{99}
\bibitem{rebunching} The experimental spectral intensity in the central part of the muti-peak region is lower compared to the theoretical values, indicating a reduction in electron rebunching of about 30\%. {P}ossible reasons for this effect may include a non-{G}aussian initial electron energy distribution, a correlated energy spread or a microbunching instability in the electron beam.
\end{thebibliography}

%

\end{document}